\documentstyle[11pt,moriond,epsfig]{article}

\def\Journal#1#2#3#4{{#1} {\bf #2}, #3 (#4)}

\def\be{\begin{equation}}
\def\ee{\end{equation}}
\def\bea{\begin{eqnarray}}
\def\eea{\end{eqnarray}}

\newcommand{\figw}[2]{%
\centerline{\includegraphics[width=#2]{#1}}%
}
\begin{document}
\newcommand{\ps}{{\sc Planck Surveyor}}
\newcommand{\pl}{{\sc Planck}}
\vspace*{4cm}
\title{CMB ANISOTROPIES ON CIRCULAR SCANS}

\author{{\bf J. Delabrouille}, R. Gispert and J.-L. Puget}

\address{Institut d'Astrophysique Spatiale, CNRS \& Universit\'e Paris XI, 
b\^at 121, 91405 Orsay Cedex, France}

\maketitle\abstracts{
We address the problem of map-making with data from the \ps\ 
High Frequency Instrument,
with an emphasis on the understanding and modelling of instrumental
effects, and in particular that of sidelobe straylight.
}

\section{CMB mapping: A complex problem}

The goal of a CMB satellite experiment as \ps\ is
a precise measurement of the temperature of the microwave sky,
$T(\vec n)$, as a function of direction $\vec n$. Data samples obtained
with the experiment, however, are not directly quantities of
interest, but rather
depend in a complex way on sky temperature, on the instrument, 
and on the scan strategy.

\subsection{Model of the measurement}

The complex process of 
radiation detection and signal production which relates the
detected signal to the temperature field on the sky
may be modelled, for a given detector, by:
\be
s(t) = h(t) \star \left [ \eta(t) \int {\rm d}\nu 
\left ( b(\nu,t) + 
H_{\rm filters}(\nu)
\int_{(4\pi)}{\rm d}\Omega L_\nu({\rm R}_t(\vec n)) T(\vec n)
\right )
\right ] + n(t)
\label{eq:measurement}
\ee
$h(t)$ is a global impulse
response of the detector and electronics, $\eta(t)$ is the
detector efficiency, 
assumed to be (possibly) slowly time dependent, 
$H_{\rm filters}(\nu)$ is the transmission of the 
filters which set the frequency band of observation, $L_\nu(\vec n)$ the 
frequency-dependent radiation pattern of the antenna (including all optical
elements), ${\rm R}_t$ a time-dependent rotation which reflects the 
scan strategy, and $T(\vec n)$ the temperature on the sky which one wishes
to measure. 
Randomness in the process of detection (in photon arrivals, in
electronic processes in the detectors) is represented by a noise term, $n(t)$.
Finally, $b(\nu,t) $ is a self-emission term, due to radiation from 
instrumental parts (filters,
slightly emissive mirrors...) impinging on the detectors.

We have written equation~\ref{eq:measurement} in terms of
continuous functions $s(t)$ and $T(\vec n)$. In the following, 
we will use sampled 
and pixelised versions $s_k$ and $T_p$ when convenient.
The problems of digitalisation and pixelisation, 
interesting by themselves, are left out of this discussion.

\subsection{Data Processing}

For optimal mapping of the CMB anisotropies, it is necessary
to optimise mission parameters and options (optimisation of $h(t)$, 
$H_{\rm filters}(\nu)$, $L_\nu(\vec n)$, ${\rm R}_t$...) 
as well as methods of data processing 
for recovering sky temperature maps from $s(t)$.
The traditionnal approach of data processing in the context of 
CMB mapping consists in modelling the measurement as 
$s_k = A_{ki} y_i + n_k$
where $s_k$ is the vector of signal samples (measurements),
$A$ a linear operator,
$y_i = (T_{\rm pixel},X_i)$ a vector of quantities to be determined
from the data set, where $T_{\rm pixel}$ are sky temperatures in a set
of pixels, and $X_i$ other quantities contributing to the signal.
$n_k$ is a vector of noise realisations.
If we assume that the noise autocorrelation matrix 
$N = \langle n \, n^T \rangle$ is known, then the inversion can
be performed by computing the estimator:
\be
\hat{y} = [A^T  N^{-1} A]^{-1} \,A^T  N^{-1} s_k
\ee
This method, in principle, produces the best unbiased estimate of the sky.
It is 
in essence the one that has been used by the DMR team \cite{janssen92}.
It suffers, however, from a few drawbacks
that make it impractical for \pl. The first is the size of the 
linear system to be manipulated (a few million pixels on the sky, and a few
billion samples per detector for a 1-year mission). The second is that
it is not clear {\it a priori} that a relevant set of parameters $X_i$ 
can be identified and the remaining noise correlation matrix estimated reliably 
so that the measurement can be meaningfully modelled by a linear equation
connecting $s_k$ to $y_i$.

We propose a way of analysing  the \pl\ data
which relies more specifically on the properties of the \pl\ scanning. 
The \pl\ satellite is spun
at $f_{\rm spin} = 1 \, {\rm rpm}$ (0.0167 Hz) around a spin axis, which is 
moved in 5 arcminute steps every 2 hours 
or so in order to follow the apparent yearly motion of the sun and keep the
spin axis roughly anti-solar. 
Because of this, the data reduction (for one detector)
can be decomposed into 2 steps:
for a given, fixed, spin-axis position, about
120 consecutive scans can be averaged into one single ring, after what
about 4000 rings (for a one year mission, for one detector)
need to be reconnected into a map of the sky. 
The first step (obtaining rings from data streams) reduces the amount of data 
samples by a factor of 120 in a non-destructive 
way and filters out efficiently all non-synchronous systematic 
effects. Scan-synchronous effects, which
can not be identified or removed in the first step, can be taken care of in 
a second step of map reconstruction from a set of rings 
(from all detectors in a channel, eventually).

\section{From data streams to rings}

\begin{figure}[htb]
  \begin{center} 
  \figw{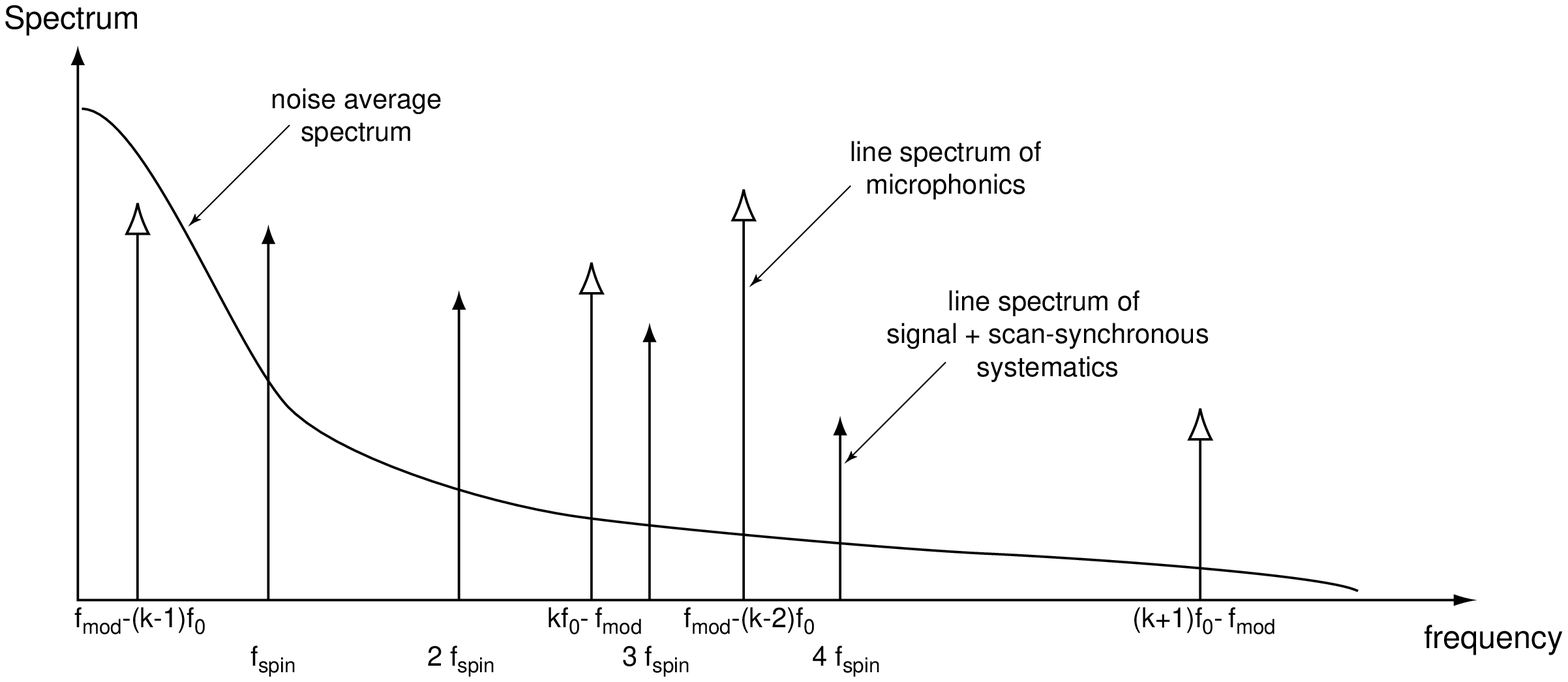}{0.7\textwidth} 
  \caption[]
  {Schematic spectra of signal, noise and systematics for periodic
  scanning, after demodulation. This spectrum is representative of 
  what is expected
  for the \pl\ HFI for one spin-axis position. Possible contribution of 
  microphonics due to periodic vibrations yields, after demodulation, a 
  non-synchronous line spectrum.
  }
  \label{fig:superpos-spectres}
  \end{center}
\end{figure}

The first step of our analysis is to build a single ring of data using
a set of $N_{\rm turns} = 120$ consecutive scans.
A plot of the expected distribution of the power of the signal as a function 
of frequency is shown on figure~\ref{fig:superpos-spectres}. Signal and 
scan-synchronous systematics have a line spectrum concentrated at harmonics
of the spinning frequency $f_{\rm spin}$. The rejection of 
non-synchronous signals is possible by building a numerical filter which keeps
harmonics of the spinning frequency and filters out other frequencies. 
Simple co-addition of samples is one such filter, but it is not necessarily 
optimal. Another option involves multiplication of the signal by a weighting
function (apodisation) prior to co-addition and computation of Fourier modes.
The filter should be adapted to the shape of the noise spectrum
and the location of non-synchronous systematic lines.
Its efficiency, whatever the method, is limited by the 
finite size of the observation window. The resolution in frequency around each
harmonic is of the order of $f_{\rm spin}/N_{\rm turns}$. 

Scan synchronous effects, or quasi synchronous noise which cannot be separated
from the signal by the filter due to insufficient resolution, cannot be 
removed from the useful astrophysical signal at this stage. Such 
signals include those induced by scan-synchronous temperature 
fluctuations of the payload, and sidelobe effects.

An additionnal use of the analysis on rings is the possibility to compare
the spectra of the cosmological signal to that of the 
noise \cite{delabrouille97}. This is a precious tool for the 
optimisation of instrument and mission parameters.

\section{Map-making from a set of rings}

Assume we now want to reconnect the set of \pl\ rings into a map.
When the main beam of the instrument comes back to a given 
pixel of the sky, we get a redundant measurement of the same useful 
astrophysical quantity in a different instrumental configuration. Such 
redundancies help identifying scan-synchronous effects reprojected on each 
ring by comparing measurements obtained at crossing points between
different rings.

As an illustration of issues 
related to the identification and subtraction of scan-synchronous effects,
we will discuss the problem of sidelobe straylight.
The measurement sample $s_k$ from the detector at a time $t_k$
can be represented, in sampled form, as
\begin{eqnarray}
s_k & = & u_k + \sum_i L_i T_{\rm sky}(p_k(i)) + n_k
\label{eq:sk1a} \\
    & = & u_k + M_{kip}L_iT_p + n_k
\label{eq:sk1b}
\end{eqnarray}
It has been assumed that the (two-dimensional, band-averaged)
sidelobe antenna pattern and the (band-averaged) sky have been pixelised 
into one dimensional 
vectors $L_i$ and $T_p$ repectively. The notation used in
equation~\ref{eq:sk1a},
supposes that for each satellite orientation 
there is an unambiguous correspondance between sky
pixels and antenna pattern pixels, represented by the map $p_k(i)$,
but this constraining assumption is not necessary when the notation of 
equation~\ref{eq:sk1b} are used. 
$u_k$ is the useful temperature fluctuation in the main beam, and $n_k$
the noise (here assumed to be white noise).

The anisotropies we want to measure do not generate significant 
sidelobe signals. Only bright features (galactic plane, dipole) contribute
significantly. A numerical estimate of the sidelobe contribution shows 
in particular that the Galaxy may generate significant (above noise levels)
sidelobe signals,
especially in the highest frequency channels.

The absolute temperature of these bright regions of the sky 
will be measured extremely precisely by \pl\ (to a relative accuracy of the
order of $10^{-2}$ to $10^{-3}$ at least) because noise
and systematic effects of any kind (after first-order correction) 
will be much lower in amplitude than the
corresponding signals (which have amplitudes of a few
millikelvin). 
Therefore, equation~\ref{eq:sk1b} can be rewritten as 
$s_k = u_k + A_{ki} L_i + n_k$,
where $A_{ki} = M_{kip}T_p$ is a $N_k \times N_i$ known matrix, which depends only on the 
absolute temperature of the sky and on the scan strategy.
A possible method to solve the system for $u_k$ and $L_i$ simultaneously
is an iterative one:
\vspace{2mm}\\
\noindent
{\bf a}- We first get an estimate of the CMB anisotropies on the sky by
standard methods (for instance
averaging measurements), neglecting sidelobe effects (or after first-order
sidelobe correction using prior knowledge of the antenna pattern). 
This yields a 
first order map $\widehat{\Delta T}(\theta, \phi)$.
\vspace{2mm}\\
\noindent
{\bf b}- An estimator $\widehat{u_k}$ of $u_k$ is computed for each sample $k$ by 
$\widehat{u_k} = \Delta T_1(\theta_k, \phi_k)$, where $(\theta_k, \phi_k)$ 
defines the direction of pointing of the main beam for sample $k$.
Differences $\widehat{\delta_k} = s_k - \widehat{u_k}$ are computed.
\vspace{2mm}\\
\noindent
{\bf c}- The system $\widehat{\delta_k} = A_{ki} L_i$ is solved
for an estimator $\widehat{L_i}$ of $L_i$. If the noise is white, the 
solution is simply $L_i = [A^T A]^{-1} \, A^T \widehat{\delta_k}$.
\vspace{2mm}\\
\noindent
{\bf d}- The process is iterated with a new estimate of $\delta_k$
obtained by correcting for the estimated sidelobe effects using the 
estimator of $L_i$.
\vspace{2mm}\\
There are two important conditions for this prescription to work in practice.
The first is that the size
of the system be small enough that several iterations can be effectively 
computed. The second is that matrix $[A^T A]$ be regular.
The fulfillment of the first condition is insured by the low spatial 
frequency character of the sources of sidelobe signal, which
have significant angular sizes of at least one degree, but usually much more
(galactic emission, CMB dipole). Using
a model of the sky obtained from extrapolations of the DIRBE data and the
DMR dipole templates in the \ps\ wavebands, and several models of \ps\
sidelobes obtained by numerical simulations, we have shown that
there is no significant
difference between convolution signals calculated using half a degree and 
three degree resolution maps. It is therefore not necessary, for sidelobe 
correction, to pixelise the lobe
with pixels smaller than a few degrees. The size of vector $L_i$ is then
a few thousand entries only. 
The fulfillment of the second condition has been checked with the use of 
numerical simulations. 
For the \pl\ scan strategy, using as a sky template the galactic emission
as measured by DIRBE, we have shown that matrix $[A^T A]$ is indeed
regular, which means that the galactic plane can be used as a source to map 
the sidelobes. This is very useful in particular for the highest frequency 
channels of \pl\, where the sidelobe contamination from galactic dust
is the largest.

\section{Conclusion}

We have shown that the scan strategy of \ps\ along rings on the sky
allows to decompose the 
problem of converting data streams into CMB anisotropy maps in two
independent steps. This makes the problem tractable numerically, and helps 
in analysing and monitoring the impact of systematic effects. In particular,
a promising method for the identification and removal of 
sidelobe signals has been 
developped.



\end{document}